\documentclass[twocolumn,prb,showpacs]{revtex4}
\usepackage{bm}
\usepackage{dcolumn}
\usepackage{graphicx}
\begin{document}

\title{Two-dimensional structure in a generic model of triangular proteins
and protein trimers}

\author{Philip J. Camp}

\email[E-mail:]{philip.camp@ed.ac.uk}

\author{Peter D. Duncan}

\affiliation{School of Chemistry, University of Edinburgh, West Mains
Road, Edinburgh EH9 3JJ, United Kingdom}

\date{\today}


\begin{abstract}
Motivated by the diversity and complexity of two-dimensional crystals
formed by triangular proteins and protein trimers, we have investigated
the structures and phase behavior of hard-disk trimers. In order to mimic
specific binding interactions, each trimer possesses on `attractive' disk
which can interact with similar disks on other trimers {\it via} an
attractive square-well potential. At low density and low temperature, the
fluid phase mainly consists of tetramers, pentamers, or hexamers. Hexamers
provide the structural motif for a high-density, low-temperature periodic
solid phase, but we also identify a metastable periodic structure based on
a tetramer motif. At high density there is a transition between
orientationally ordered and disordered solid phases. The connections
between simulated structures and those of 2D protein crystals -- as seen
in electron microscopy -- are briefly discussed.
\end{abstract}

\maketitle

\section{Introduction}
\label{sec:introduction}

Two-dimensional (2D) materials present some fascinating challenges to
condensed-matter theory, with even the most simple 2D systems harboring
surprises. One of the most famous problems involves the precise
description of melting in 2D solids made up of hard, disk-like particles
with short-range repulsive
interactions.\cite{Kosterlitz:1973/a,Nelson:1979/a,Young:1979/a,Binder:2002/a}
Specifically, does the fluid undergo a weak first-order transition to the
solid, or is there an intermediate hexatic phase linked by two continuous
phase transitions? Related avenues of research concern the existence of
exotic phases in systems made up of more complex particles, such as
(non)periodic solids of hard-disk
dimers,\cite{Wojciechowski:1991/a,Wojciechowski:1992/a} pentamers and
hexamers,\cite{Wojciechowski:2003/a} tetratic phases of hard
squares\cite{Wojciechowski:2004/a} and hard rectangles,\cite{Donev:2005/a}
and orientationally ordered solids of hard pentagons and
heptagons.\cite{Schilling:2005/a} The effects of additional interactions
on the phase behavior and dynamics of 2D systems are also of interest, as
evidenced by recent studies on dipolar potentials in the context of
magnetic colloids.\cite{Weis:2003/a} Such models provide an ideal testing
ground for condensed-matter theories, and in some cases challenge our most
fundamental understanding of the properties of matter.

Despite their simplicity, 2D models can provide reliable descriptions of
some real, and rather complex, experimental situations. For example, in a
number of recent studies, 2D models have been employed to help interpret
and understand the clustering and crystallization of proteins at
interfaces. The conformations and interactions of proteins are central to
biological activity, and ideally one would like to investigate these
properties {\it in vivo}. Unfortunately, structural information is most
commonly obtained from X-ray diffraction studies on crystals. There is a
class of proteins, however, that can be studied under conditions
resembling those {\it in vivo}. Membrane proteins constitute a large class
of molecules found within the lipid bilayers that constitute cell walls.
They fulfill a variety of roles, such as controlling the selective
transport of ions and molecules across cell membranes, or providing
binding sites for other molecules on to the membrane. The structures of
membrane proteins can be studied by deposition on to a surface, alongside
lipids, to form either low surface-coverages or 2D crystals; the
hydrophobic lipids help to mimic the interior of the membrane. Electron
microscopy or atomic-force microscopy can then be used to image directly
the clustering and packing of proteins at the solid-air
interface.\cite{Werten:2002/a} In many cases, the ordering of proteins can
be rationalized on the basis of their gross shapes (the way in which those
shapes would `tile the plane') and the presence of specific binding
interactions between domains on different molecules. For example, the
surface structure of bacteriorhodopsin (a transmembrane protein) is
comprised of a close-packed array of trimers, each made up of monomers
that resemble $120^\circ$ sectors of a circle.  Monte Carlo (MC)
simulations of hard sectors -- with an additional attractive square-well
potential to mimic specific binding interactions -- yield insight on the
self-assembly and subsequent crystallization
processes.\cite{Jagannathan:2002/a} In another application, the ordering
in 2D crystals of annexin V -- another `triangular' membrane protein --
was reproduced in simulations of a hard-disk model decorated with an
appropriate orientation-dependent potential to mimic the locations of the
specific binding sites on the protein. Experimentally observed honeycomb
and triangular structures were captured by the molecular model. These
examples show that the basic physics of large-scale structural order in 2D
protein crystals can be studied with simple models, and without resorting
to atomistically detailed -- and hence very expensive -- computer
simulations.

There are a large number of proteins which are either inherently
triangular in shape, or otherwise form trimeric
structures.\cite{Hsu:1997/a,Wang:2000/a,Yin:2000/a,Conroy:2004/a,%
Gibbons:2004/a,Govaerts:2004/a,Kitmotto:2005/a} For example, rotavirus
inner capsid protein V6 forms trimers resembling equililateral triangles,
which pack in 2D crystals (space group $p6$).\cite{Hsu:1997/a} Specific
fragments of prion proteins found {\it in vivo} form trimeric units that
crystallize in to a 2D structure (space group $p3$).\cite{Govaerts:2004/a}
Finally, we highlight an example in which a membrane fusion protein (from
the Semiliki forest virus) is seen to form pentagons of trimers, with the
center of the pentagon raised slightly out of the
plane.\cite{Gibbons:2004/a} Some semblance of local five-fold coordination
can also been seen in TetA -- a roughly triangular transporter protein --
at moderate surface coverages.\cite{Yin:2000/a}

Motivated by the diversity of 2D crystal structures exhibited by trimeric
protein units, and also by the observation of five-fold
coordination,\cite{Gibbons:2004/a} we have investigated the structure and
phase behavior of model trimeric molecules made up of hard disks.  In
order to mimic specific binding interactions, such as those that might
give rise to local five-fold coordination, we focus on an equililateral
triangle of three hard disks at contact, in which one disk can interact
with the corresponding disks on other molecules {\it via} a short-range
attractive square-well potential. As we will show below, this raises the
possibility of generating orientational order within simple close-packed
structures, and also offers the opportunity of forming clusters at low
surface coverages. Using MC simulations, we map out the phase diagram of
the model system, and characterize the structures of the low-density
clustered fluid and high-density solids which are formed at low
temperature. The remainder of the article is organized as follows. In
Section \ref{sec:model} we describe the molecular model, and summarize the
simulation methods. The results are presented in Section
\ref{sec:results}, and Section \ref{sec:discussion} concludes the paper.

\section{Model and methods}
\label{sec:model}

The molecular model consists of three hard disks, each of diameter
$\sigma$, fused at mutual contact to form an equilateral triangle. Two of
the disks on each molecule are purely repulsive, and interact with all
other disks in the system through the potential
\begin{equation}
u(r) = \left\{
       \begin{array}{ll}
        \infty & r < \sigma \\
        0      & r \geq \sigma
       \end{array}
       \right.
\label{eqn:urep}
\end{equation}
where $r$ is the separation between the centers of two disks. The third
disk on each molecule carries a central attractive interaction site; these
`attractive' disks interact with each other {\it via} the potential
\begin{equation}
u(r) = \left\{
       \begin{array}{ll}
        \infty    & r < \sigma \\
        -\epsilon & \sigma \leq r < \lambda\sigma  \\
        0         & r \geq \lambda\sigma 
       \end{array}
       \right.
\label{eqn:uatt}   
\end{equation}
where $\lambda > 1$ controls the range of the attraction. This potential
crudely mimics an effective attraction between vertices of the molecular
triangles, which might arise through specific interactions (e.g. hydrogen
bonding, disulfide bridges, effective solvophobic interactions).  

The parameter $\lambda$ will clearly have a crucial role to play in the
thermodynamics of the system. If $\lambda >> 1$ then one should anticipate
a conventional phase diagram containing a vapor-liquid transition, and a
fluid-solid transition. The orientation of a trimer can be defined by a
vector ${\bf n}$ joining the geometrical center of the trimer with the
center of the attractive disk. It is unlikely that there would be any
periodic orientational ordering of ${\bf n}$ in the solid phase; if two
trimers can interact favorably irrespective of the mutual orientation,
then on entropy grounds the orientations will be disordered. In the
opposite extreme, $\lambda \agt 1$, the molecules will feel the
orientation dependence of the net trimer-trimer potential, and ultimately
we might expect the vapor-liquid transition to disappear from the
equilibrium phase diagram. Indeed, in a pure square-well hard-sphere
fluid, condensation becomes metastable with respect to freezing when
$\lambda < 1.25$.\cite{Pagan:2005/a} In the present case, an interaction
range $\lambda < \sqrt{3}$ guarantees that attractive sites must face each
other directly in order to interact; when $\lambda > \sqrt{3}$ it is
possible for an attractive disk to be within interaction range of a trimer
even if it approaches from `behind'. With these comments in mind, we have
chosen to study a system with $\lambda = 1.25$. The ratio of
$\lambda\sigma$ to the (angle-averaged) diameter of the trimer is smaller
than that in a pure square-well hard-sphere system with the same value of
$\lambda$, and assuming some sort of correspondence between two and
three-dimensional systems, we do not anticipate there being a vapor-liquid
transition in the equilibrium phase diagram. On the other hand, because
the trimers have to attain quite specific mutual orientations in order to
interact favorably (since $\lambda < \sqrt{3}$), we should expect to see
some sort of non-trivial structure in fluid and solid phases at low
temperatures.

Systems of $N=120$ trimers were studied using MC simulations either in the
isothermal-isobaric ($NpT$) ensemble or the canonical ($NAT$)  
ensemble.\cite{Allen:1987/a} The simulation cell was rectangular with
dimensions $L_{x}$ and $L_{y}$, and area $A=L_{x}L_{y}$. Each MC cycle
consisted of one translational trial move and one orientational trial move
for each of $N$ randomly selected molecules. To help equilibrate dense
phases, every fifth MC cycle included $N$ trial moves in which a randomly
selected trimer was rotated by $\pm 120^\circ$. In $NpT$ simulations of
solid phases, $L_{x}$ and $L_{y}$ were varied independently; in $NpT$
simulations of fluid phases, the simulation cell was constrained to be
square. For most thermodynamic state points typical equilibration runs
consisted of $\sim 10^{5}$ MC cycles, but some points (close to phase
transitions) required $\sim 10^{6}$ MC cycles. Production runs were
typically $\sim 10^{5}$ MC cycles. We define the following dimensionless
units in terms of the square-well depth, $\epsilon$, and the hard-disk
diameter, $\sigma$: number density $\rho^{*}=N\sigma^2/A$; temperature
$T^{*}=k_{B}T/\epsilon$; pressure $p^{*}=p\sigma^{2}/k_{B}T$.

\section{Results}
\label{sec:results}

\begin{figure}[tb] \includegraphics[scale=0.33]{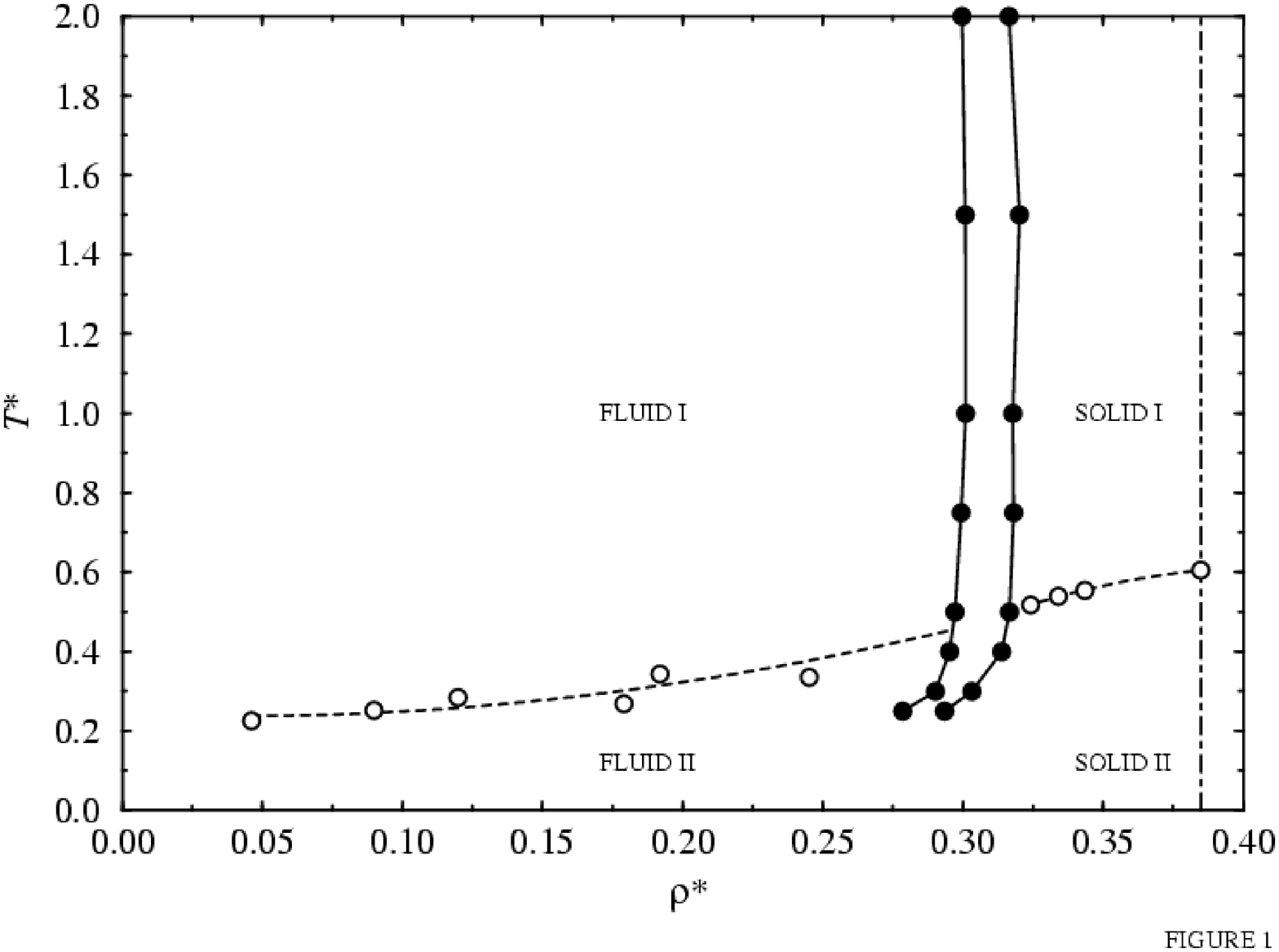}
\caption{\label{fig:phasediag} Phase diagram of the model trimer system in
the density-temperature ($\rho^{*}$-$T^{*}$) plane: (solid points and
solid lines) approximate fluid-solid phase boundaries, assumed to be first
order; (open points and dashed lines) boundaries between high-temperature
unclustered states and low-temperature clustered states, as evidenced by
maxima in the heat capacity along isobars; (dot-dashed line) close-packed
density, $\rho_{\rm cp}^{*}=2/3\sqrt{3}\simeq 0.3849$.} \end{figure}

\begin{figure*}[tb]\includegraphics[scale=0.96]{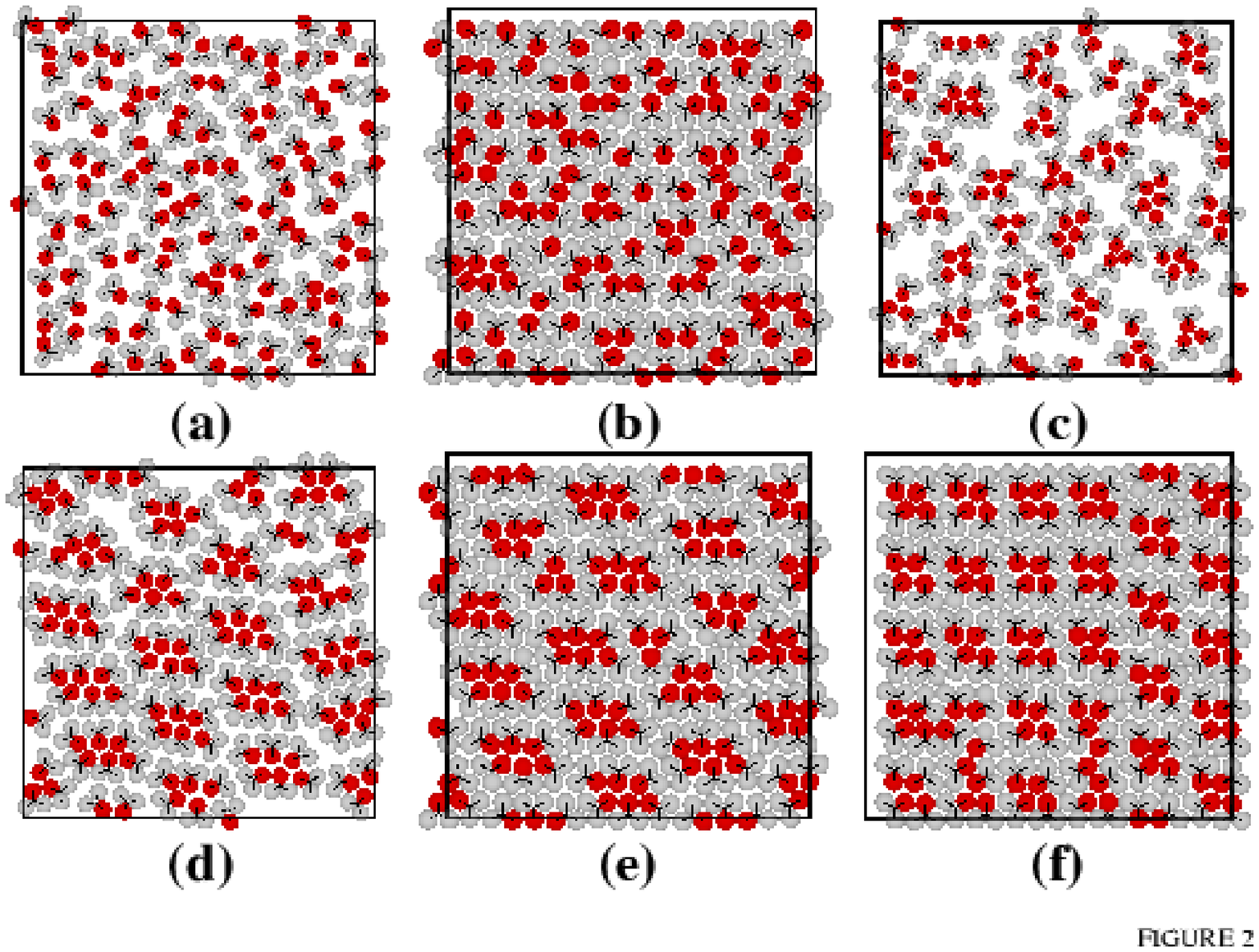}
\caption{\label{fig:snapshot} (Color online) Configuration snapshots from
$NpT$ simulations: (a) normal fluid phase (fluid I) at $T^{*}=2$,
$p^{*}=2.5$, $\rho^{*}=0.259$; (b) orientationally disordered $AB$-solid
phase (solid I) at $T^{*}=2$, $p^{*}=12$, $\rho^{*}=0.345$; (c) clustered
fluid phase (fluid II) at $T^{*}=0.25$, $p^{*}=0.75$, $\rho^{*}=0.222$;
(d) metastable state at $T^{*}=0.25$, $p^{*}=2.6$, $\rho^{*}=0.290$; (e)
orientationally ordered $AB$-solid phase (solid II) at $T^{*}=0.25$,
$p^{*}=12$, $\rho^{*}=0.349$; (f) metastable orientationally ordered
$AA$-solid at $T^{*}=0.25$, $p^{*}=20$, $\rho^{*}=0.356$. In each case the
attractive disks are colored dark gray (red online), the repulsive disks
are colored light gray, and all disks are drawn with diameter $1\sigma$.}
\end{figure*}

The phase diagram of the model trimers in the density-temperature
($\rho^{*}$-$T^{*}$) plane is sketched in Fig.~\ref{fig:phasediag}. Before
detailing the determination of the phase boundaries, the characteristics
of the different phases will be described. There are four distinct regions
in the phase diagram. At low density and high temperature, a normal fluid
phase is in evidence (fluid I). A typical simulation configuration is
shown in Fig.~\ref{fig:snapshot}(a). There is neither translational nor
long-range orientational order in the system.

At high density and high temperature, the stable solid phase (solid I)
possesses an orientationally disordered structure (in the sense that ${\bf
n}$ is disordered) with the trimers close-packed to form alternating rows
displaced by $\sigma/2$. Figure \ref{fig:snapshot}(b) shows both the lack
of orientational order, and the registry between alternating rows. Notice
the black bonds showing how the disks are connected within the trimers; we
call this an `$AB$' structure to denote the alternating alignment of the
rows. The close-packed rows resemble those formed by VP6,\cite{Hsu:1997/a}
although the registry between the rows is different. At the end of this
section, we will briefly discuss the possibility of solids with other
close-packed structures.

\begin{figure}[tb] \includegraphics[scale=0.33]{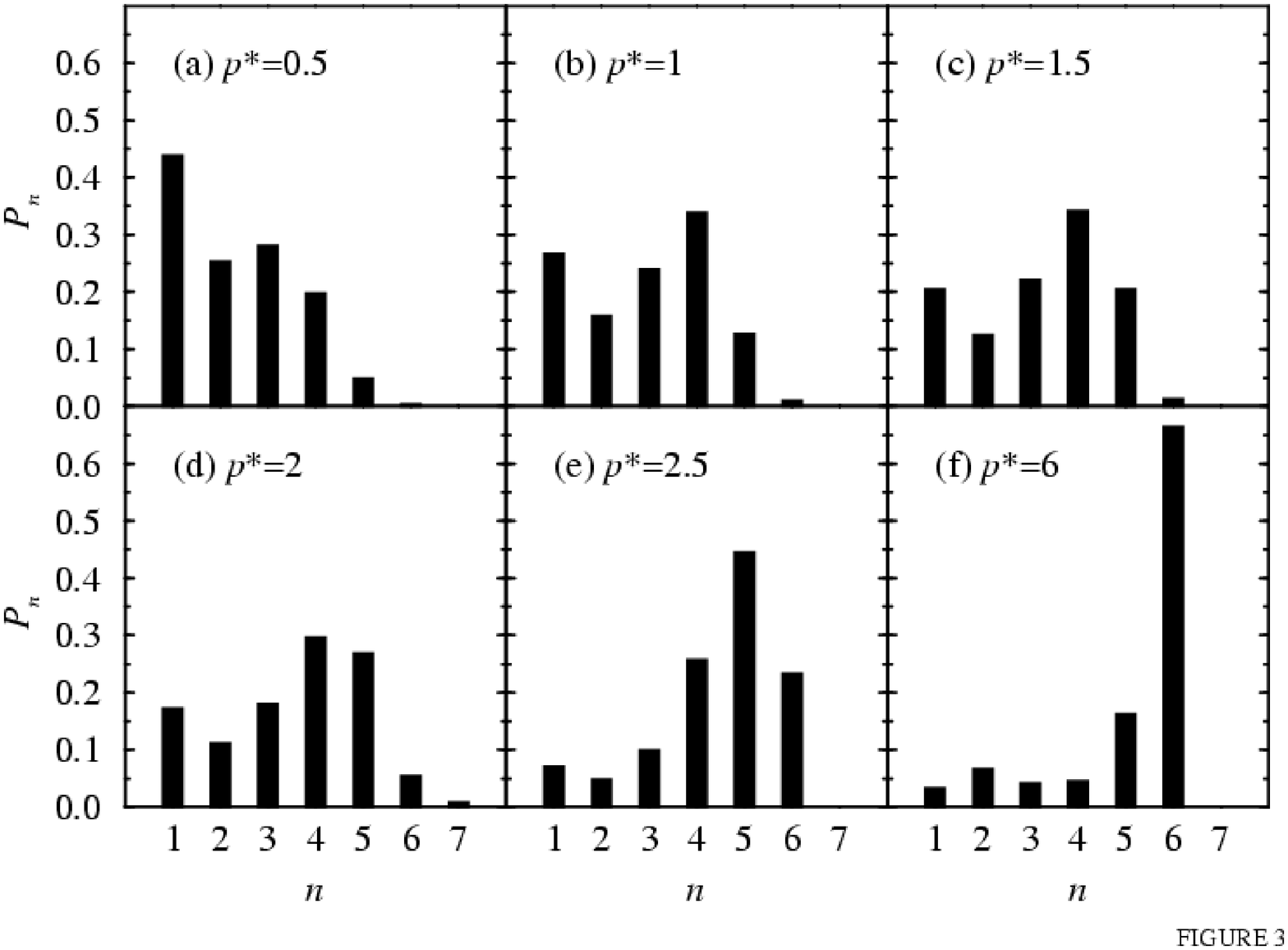}
\caption{\label{fig:cluster} Cluster distributions for systems along the
isotherm $T^{*}=0.3$: (a) $p^{*}=0.5$, $\rho^{*}=0.188$; (b) $p^{*}=1$,
$\rho^{*}=0.230$; (c) $p^{*}=1.5$, $\rho^{*}=0.252$; (d) $p^{*}=2$,
$\rho^{*}=0.265$; (e) $p^{*}=2.5$, $\rho^{*}=0.280$; (f) $p^{*}=6$,
$\rho^{*}=0.329$. In (a)-(e) the system is fluid, whilst in (f) the system
is solid (II).} \end{figure}

At low temperature and low density we find a highly associated fluid
(fluid II), in which the attractive disks aggregate to form distinct
clusters. A typical configuration is shown in Fig.~\ref{fig:snapshot}(c),
which exhibits a broad distribution of cluster sizes. To identify
clusters, we employ the obvious criterion that two trimers with attractive
disks within interaction range belong to the same cluster. With this
definition in mind, Fig.~\ref{fig:snapshot}(c) shows that, in general, the
attractive disks within the clusters form close-packed motifs, rather than
loose arrangements of disks on the circumference of a ring.  For clusters
of three trimers there is no distinction, whereas for four or more trimers
the close-packed arrangement is more favorable; in a ring, each disk would
have two nearest neighbors, whereas close-packed motifs can accommodate
more than two direct contacts. In Fig.~\ref{fig:cluster} we show the
probability distribution function of clusters containing $n$ molecules, at
different pressures along an isotherm with $T^{*}=0.3$. As the pressure
and density increase, the distributions show peaks at progressively higher
values of $n$. At the highest fluid-density shown -- $\rho^{*}=0.280$,
Fig.~\ref{fig:cluster}(e) -- the most probable cluster size is $n=5$. We
had hoped that these clusters would adopt a pentagonal structure, but
instead the attractive disks form `Olympic rings' motifs, such as those
shown in Fig.~\ref{fig:snapshot}(c). The maximum disk-disk separation in a
perfect pentagon of disks is $\sqrt{2(1-\cos{108^\circ})}\sigma \simeq
1.62\sigma$, which is longer than the range of the potential studied in
this work. Hence, to minimize the energy, the cluster will contract to
form a close-packed structure. Perhaps pentagonal clusters would be formed
in a system with $1.62 \leq \lambda \leq \sqrt{3}$? (The upper limit means
that there can be no other disks between two interacting attractive
disks.) We did some test runs in the fluid phase with $\lambda = 1.7$, but
no pentagonal clusters were observed. If anything, fewer distinct clusters
were in evidence as compared to $\lambda = 1.25$, presumably because it is
less crucial that the trimers attain a specific mutual orientation in
order to interact.

Upon compression of the low-temperature fluid we often encountered
metastable structures, such as that shown in Fig.~\ref{fig:snapshot}(d).
This clearly shows a predominance of $n=6$ clusters, with the attractive
disks close packed to form a parallelogram motif, but the clusters are not
yet fully packed in to a solid structure. This process is completed upon
further compression, to form a $p2$ periodic solid (solid II), a defective
example of which is shown in Fig.~\ref{fig:snapshot}(e). In simulations of
the high-density solid II phase, the initial configuration consisted of
the appropriate $AB$ structure, but with ${\bf n}$ for each molecule
chosen randomly from the three molecular arms; the orientational structure
shown in Fig.~\ref{fig:snapshot}(e) develops spontaneously. The cluster
distribution for such a solid at temperature $T^{*}=0.3$ and density
$\rho^{*}=0.329$ is shown in Fig.~\ref{fig:cluster}(f). The primary peak
is at $n=6$, but the presence of defects -- such as those shown in
Fig.~\ref{fig:snapshot}(e) -- gives rise to smaller `clusters' of
attractive disks.

\begin{figure}[tb] \includegraphics[scale=0.33]{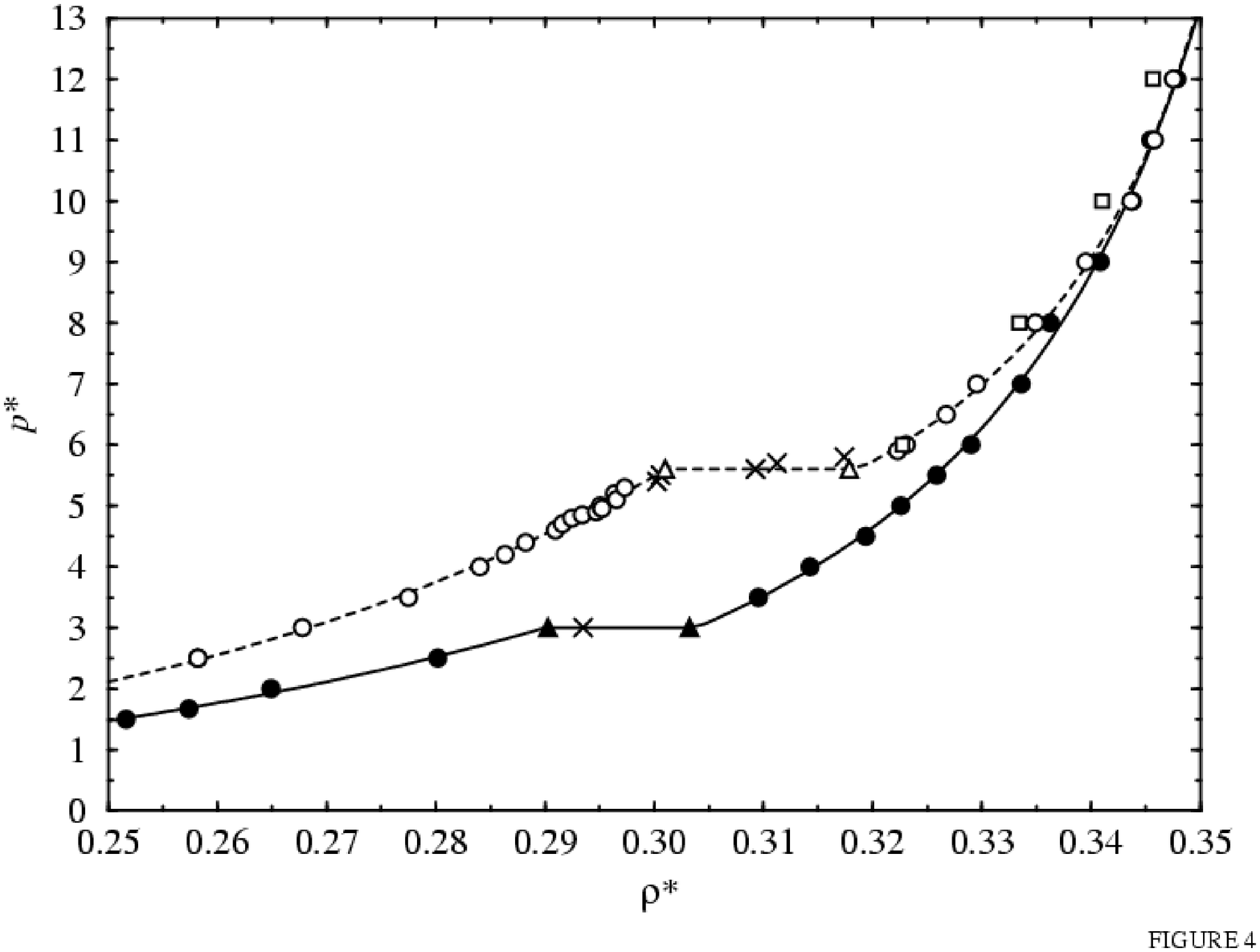}
\caption{\label{fig:eos} Equations of state along isotherms with
$T^{*}=0.3$ (solid symbols, solid lines), and $T^{*}=1$ (open symbols,
dashed lines): (circles) state points from $NpT$ simulations, with $AB$
solid phases; (squares) state points from $NpT$ simulations, with $AA$
solid phases ($T^{*}=1$ only); (crosses)  putative metastable state
points;  (triangles) approximate coexistence densities; (lines) fits to
the fluid and solid branches (see text). The statistical errors in the
$NpT$ simulation points are smaller than the symbols.} \end{figure}

The fluid-solid phase boundaries were located by monitoring the equation
of state $p(\rho)$ along selected isotherms in $NpT$ simulations.  For
each isotherm, two sets of simulations were performed: a compression
branch, starting from a low-density fluid configuration; and an expansion
branch, starting from the perfect solid structure corresponding to that
found in the compression branch at high pressure. Portions of two
representative examples ($T^{*}=0.3$ and $T^{*}=1$) are shown in
Fig.~\ref{fig:eos}. Of course, the fluid equations of state extend to much
lower densities, but these exhibit entirely conventional behavior and
hence are not shown; in particular, there is no sign of a `van der Waals'
loop which would indicate a vapor-liquid phase transition. The main
features of interest are the apparent discontinuities in the density at
what are assumed to be first-order phase transitions (we will not open up
the can of worms associated with the precise nature of two-dimensional
melting and freezing\cite{Kosterlitz:1973/a,Nelson:1979/a,Young:1979/a,%
Binder:2002/a}). In Fig.~\ref{fig:eos} we indicate distinct fluid and
solid branches in the equations of state, a number of putative metastable
states (as discussed above), and approximate tie-lines connecting the
fluid and solid coexistence densities, obtained as follows.  The fluid
branch was fitted with a virial expansion containing terms up to
$\rho^{5}$, i.e., $p/k_{B}T=\rho+\sum_{n=2}^{5}B_{n}\rho^{n}$, while the
solid branch was found to be fitted rather well by a simple van der Waals
equation\cite{Daanoun:1994/a} of the form
$p/kT=a\rho/(1-b\rho)-c\rho^{2}$, which contains a free-volume term
arising from repulsive interactions, and a mean-field term arising from
the attractions. The coexistence densities were then estimated by
extrapolating the fitted branches of the equation of state to a pressure
half way between those in the highest-density stable fluid and the
lowest-density stable solid; the metastable states were identified as
those that did not fit on to either branch and/or for which the simulation
configuration was clearly neither pure solid nor pure fluid, e.g.  
Fig.~\ref{fig:snapshot}(d). Obviously this approach provides only very
rough locations for the phase boundaries shown in
Fig.~\ref{fig:phasediag}, but some general trends are nonetheless
apparent. At very low temperatures, the coexistence densities decrease as
the system is cooled, and the transition appears to be getting weaker. At
high temperatures ($T \geq 1$) the fluid coexistence density
($\rho^{*}\simeq 0.30$) is very similar to the density at which the pure
hard-disk fluid undergoes its transition, either to a hexatic or a solid
(disk density $\rho^{*}=0.899$,\cite{Binder:2002/a} `trimer' density
$\rho^{*}=0.300$). The apparent trimer solid coexistence density
($\rho^{*}\simeq 0.32$) is significantly larger than the melting density
of hard disks (disk density $\rho^{*}\simeq 0.914$,\cite{Binder:2002/a}
`trimer' density $\rho^{*}\simeq 0.305$).

\begin{figure}[tb] \includegraphics[scale=0.33]{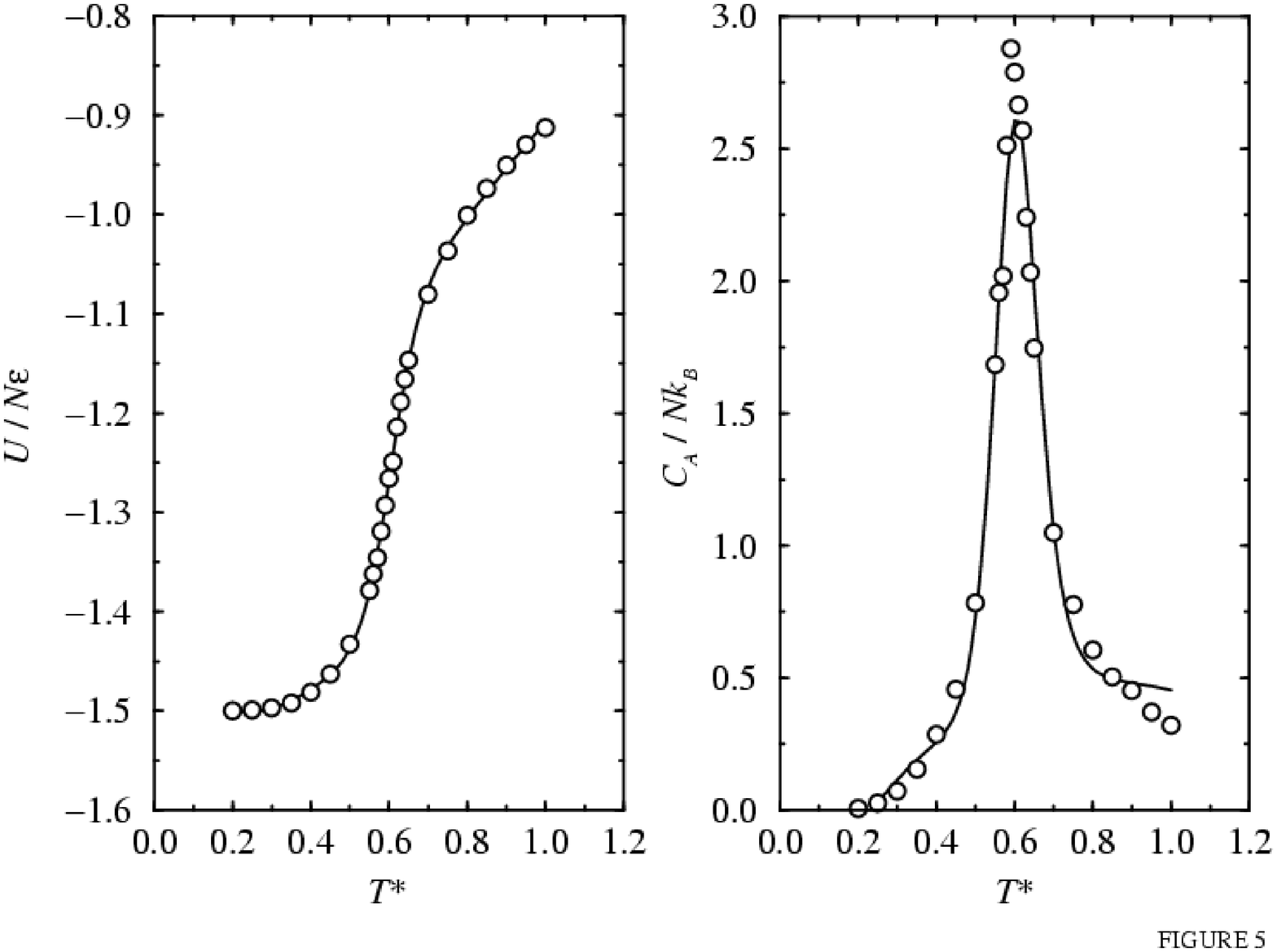}
\caption{\label{fig:heatcap} Configurational energy $U$ (left) and excess
heat capacity $C_{A}$ (right) as functions of reduced temperature $T^{*}$
at the close-packed density $\rho^{*}=2/3\sqrt{3}\simeq 0.3849$:  
(circles) simulation results; (lines) results derived from a Pad{\'e}
[5,5] fit (see text).} \end{figure}

The final piece of the equilibrium phase diagram concerns the crossover
from high-temperature orientationally disordered states to low-temperature
states that possess structural motifs arising from the clustering of the
attractive disks. To delineate the boundary between these two regimes, we
calculated the heat capacity appropriate to the statistical mechanical
ensemble being sampled. In general we used $NpT$ simulations to measure
$C_{p}=(\partial H/\partial T)_{p}$ -- where $H=U+pA$ is the enthalpy
(minus the kinetic contribution) -- as a function of temperature along an
isobar. Since clustering must be accompanied by a drop in the
configurational energy, and enthalpy, a peak in $C_{p}$ would seem to be
an obvious signal of a crossover from unclustered to clustered states. In
simulations we evaluated the usual fluctuation formula, $C_{p}=[\langle
H^{2} \rangle - \langle H \rangle^{2}]/k_{B}T^{2}$, and, as a check,
differentiated an $[n,n]$ Pad{\'e} approximant fitted to the enthalpy as a
function of $T$;
\begin{equation}
H = \frac{a_{0}+a_{1}T+a_{2}T^{2}+\ldots+a_{n}T^{n}}
         {    1+b_{1}T+b_{2}T^{2}+\ldots+b_{n}T^{n}}.
\label{eqn:pade}
\end{equation}
These two approaches yielded consistent results, and the peak in $C_{p}$
was easy to locate accurately. In general the peak height is less
pronounced at high densities, mainly due to the fact that even in the
high-temperature phase there must be some attractive disks within
interaction range due to the confinement. Thus, the most difficult
situation obtains at close packing of the trimers, $\rho_{\rm
cp}^{*}=2/3\sqrt{3}$. In this case we studied a perfect close-packed $AB$
solid, and carried out $NAT$ MC simulations with $\pm 120^{\circ}$
rotations only. We show results for the configurational energy, $U$, and
the excess constant-area heat capacity, $C_{A}=(\partial U/\partial
T)_{A}$, in Fig.~\ref{fig:heatcap}. A [5,5] Pad{\'e} fit provides a
reliable description of the energy, and the corresponding results for
$C_{A}$ are consistent with those obtained {\it via} the fluctuation
formula.

In Fig.~\ref{fig:phasediag} we show the positions of the maxima in $C_{p}$
-- and $C_{A}$ at $\rho_{\rm cp}^{*}=2/3\sqrt{3}$ -- along with separate
cubic fits to the points in the fluid and solid regions of the phase
diagram. It appears that the two branches would meet up somewhere in the
fluid-solid coexistence region. We stress that the boundaries indicated do
not represent thermodynamic phase transitions; rather, they separate
different regimes of trimer association.

Finally, we briefly consider the possibility of the trimer system adopting
other solid structures, such as the $p2$ $AA$ structure shown in
Fig.~\ref{fig:snapshot}(f), in which the close-packed (horizontal) rows
are matched with the neighboring rows. In this case, the low-temperature,
orientationally ordered solid exhibits rhombic cluster-motifs containing
only four attractive disks. Out of those four disks, two are interacting
with two other disks, and two are interacting with three other disks.
Hence, the minimum configurational energy for an $AA$ solid is
$-\frac{5}{4}\epsilon$ per trimer. In the $AB$ structure, there are six
attractive disks per parallelogram motif, of which two have two neighbors,
two have three neighbors, and two have four neighbors, giving a minimum
energy of $-\frac{3}{2}\epsilon$ per trimer. Hence, on energetic grounds,
we should expect the $AB$ structure to be thermodynamically favored. Even
at high temperature, the $AA$ structure appears to be less stable with
respect to the $AB$ structure. As an example, in Fig.~\ref{fig:eos}, we
show an $AA$-solid branch of the equation of state at $T^{*}=1$, alongside
the $AB$-solid branch. For a given pressure, the $AB$ solid has the higher
density which makes this state at least mechanically stable with respect
to $AA$. Indeed, we only ever observed the fluid spontaneously freezing in
to an $AB$ structure. Although we have not performed free-energy
calculations, it would be very surprising if an entropic effect could
compensate for the relative energetic and mechanical stability of the $AB$
phase with respect to the $AA$ phase.

\begin{figure}[tb] \includegraphics[scale=0.33]{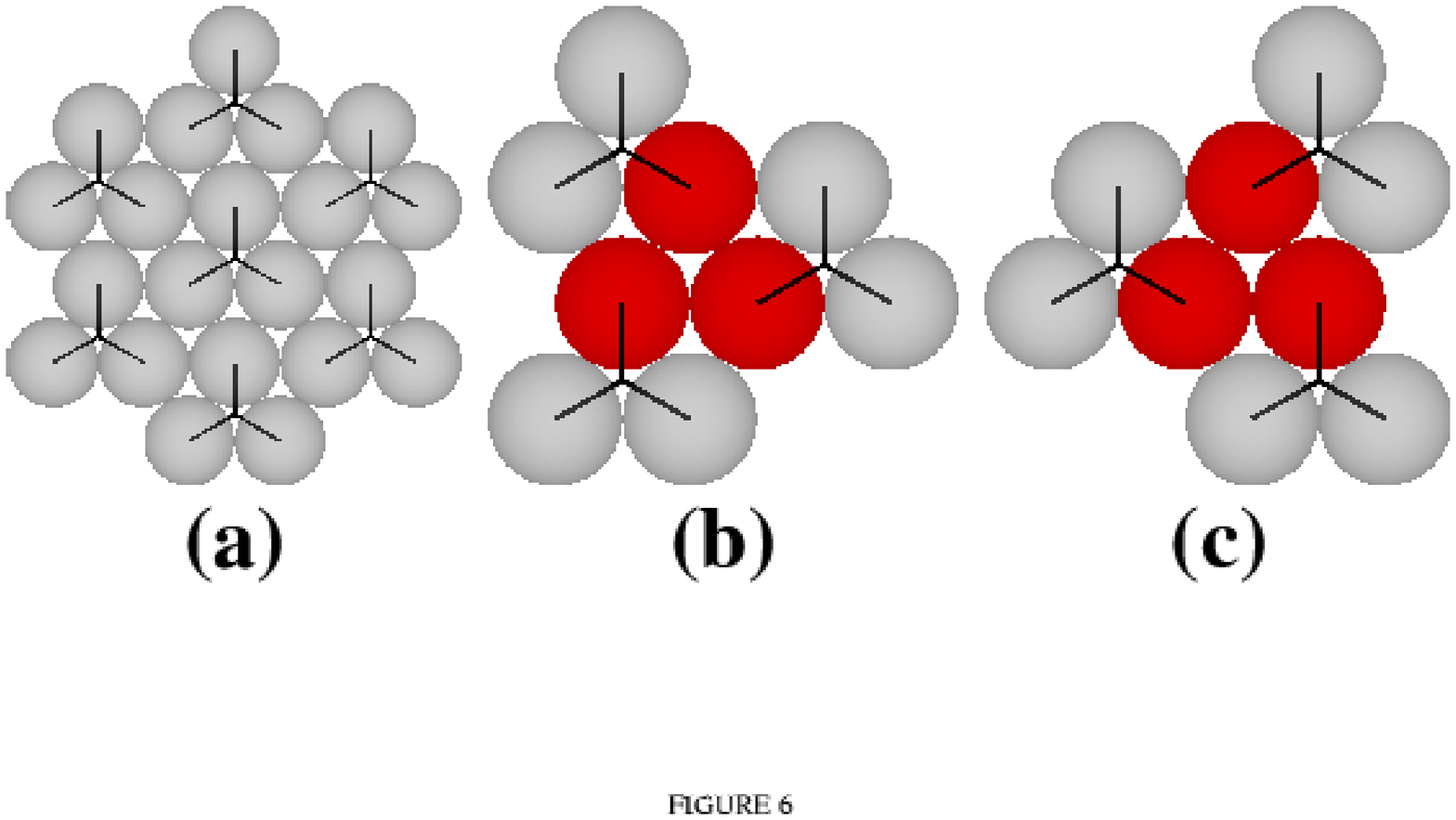}
\caption{\label{fig:hexagonal} (Color online) Illustrations of an
alternative close-packed structure: (a) without an assignment of
attractive disks; (b) and (c) mirror images of a possible structural motif
for a periodic arrangement of attractive disks. The attractive disks are
colored dark gray (red online), the repulsive disks are colored light
gray, and all disks are drawn with diameter $1\sigma$.} \end{figure}

Another possible close-packed structure is illustrated in
Fig.~\ref{fig:hexagonal}(a), without any indication of the attractive
disks. This structure resembles that adopted by 2D crystals of
TetA,\cite{Yin:2000/a} although we never saw this packing structure emerge
from our simulations. As far as our model is concerned, the absence of
this structure at low temperature is easy to understand. In
Figs.~\ref{fig:hexagonal}(b) and \ref{fig:hexagonal}(c) we illustrate
mirror images of the most obvious periodic arrangement of the attractive
disks (space group $p3$). The energy per trimer is only $-1\epsilon$, and
so this is not competitive with the $AB$ structure that is seen to emerge
spontaneously in our simulations. Free-energy calculations would be of
interest, particularly at high temperatures where entropy is everything!

\section{Discussion}
\label{sec:discussion}

In this article we have described the structure and phase behavior of a
generic model of trimeric molecules, largely motivated by recent
experimental 2D microscopy studies of clustering and crystallization in
triangular proteins and protein trimers. The molecular model consists of a
triangle of hard disks, with one of the disks participating in attractive
square-well interactions with similar disks on other trimers. The range of
the square-well potential, $\lambda\sigma$, was $1.25$ times the disk
diameter. This system crudely mimics the general shape and specific
interactions of a wide range of proteins. The model system exhibits fluid
and solid phases which, at low temperatures, possess interesting
structural motifs arising from the clustering of the `attractive' disks.

In the fluid, a distribution of clusters is in evidence, including
tetramers, pentamers, and hexamers (of trimers). In the pentamers and
hexamers, the attractive disks close-pack to form `Olympic rings' and
parallelogram shapes, respectively. We had hoped to find more open
pentagonal clusters of trimers, such as those reported in
Ref.~\onlinecite{Gibbons:2004/a}. To investigate the formation of such
clusters further, it might be interesting to study a system of hard
isosceles triangles with the unique angle equal to $72^\circ$, and a
short-range attraction operating between the corresponding vertices.

In the low-temperature solid, the basic structural motif consists of
clusters of six molecules, with the attractive disks close-packed to form
a parallelogram. A metastable solid possessing a motif made up of four
molecules was also identified. The fundamental difference between the two
situations is the registry between neighboring close-packed rows of
trimers ($AB$ versus $AA$). Even at high temperatures, the orientationally
disordered $AB$ solid is at least mechanically stable with respect to the
$AA$ solid. We identified a third structure based on hexagonal close
packing, but this structure is not competitive either, at least in terms
of energy. It would be worth performing free-energy calculations to study
these issues further.

Finally, it is worth commenting that a diverse range of 2D structures can
be generated from very simple molecular models. Fully atomistic
calculations of 2D protein structures are expensive, and, it could be
argued, yield little insight on the fundamental physics behind clustering
and crystallization. As has been shown in a variety of cases, including
the present study, the process of developing and studying simple models of
complex systems can yield some surprising results.

\begin{acknowledgments}
The provision of a studentship for PDD by the Engineering and Physical
Sciences Research Council (UK) is gratefully acknowledged.
\end{acknowledgments}


\end{document}